\documentclass{article}
\usepackage{graphics}
\usepackage{graphicx}
\title{\bf K-Essential Phantom Energy: Doomsday around the corner?}
\author{ Pedro F. Gonz\'{a}lez-D\'{\i}az.\\
Instituto de Matem\'{a}ticas y F\'{\i}sica Fundamental\\ Consejo Superior
de Investigaciones Cient\'{\i}ficas\\ Serrano 121, 28006 Madrid,
SPAIN\\ }
\date{December 1, 2003}
\begin{document}
\maketitle \large \setlength{\baselineskip}{0.9cm}

\begin{center}
{\bf Abstract}
\end{center}
In spite of its rather weird properties which include violation of
the dominant-energy condition, the requirement of superluminal
sound speed and increasing vacuum-energy density, phantom energy
has recently attracted a lot of scientific and popular interests.
In this letter it is shown that in the framework of a general
k-essence model, vacuum-phantom energy leads to a cosmological
scenario having negative sound speed and a big-rip singularity,
where the field potential also blows up, which might occur at an
almost arbitrarily near time in the future that can still be
comfortably accommodated within current observational constraints.

\vspace{.5cm}


\vspace{.5cm}


\pagebreak

The existence of phantom dark energy in the universe actually is a
possibility not excluded by observations which has recently been
widely discussed [1]. The physical properties of vacuum phantom
energy are rather weird, as they include violation of the
dominant-energy condition, $P+\rho<0$,, naive superluminal sound
speed and increasing vacuum-energy density. The latter property
ultimately leads to the emergence of a singularity - usually
referred to as {\it big rip} - in a finite time in future where
both the scale factor and the vacuum-energy density blow up [2].
The existence of a singularity in finite time was already
considered by Barrow, Galloway and Tipler in 1986 [3], even under
the much weaker conditions $\rho>0$ and $\rho+3P>0$, by assuming
that $dP/d\rho$ is not a continuous function. This can actually be
regarded as the first example of a big rip singularity. On the
other hand, if we want the weak energy condition to be preserved
one must regard the stuff of phantom energy to be made up of
axions, at least when dealing with a quintessence field [4].
Indeed, if a quintessential scalar field $\phi$ with constant
equation of state $P_{\phi}=\omega\rho_{\phi}$ is considered, then
phantom energy can be introduced by allowing violation of dominant
energy condition, $P_{\phi}+\rho_{\phi} < 0$, or what is
equivalent, rotation of $\phi$ to imaginary values,
$\phi\rightarrow i\Phi$, in the Lorentzian manifold [Notice that
if the pressure and the energy density are respectively defined as
$P_{\phi}=\frac{1}{2}\dot{\phi}^2-V(\phi)$ and
$\rho_{\phi}=\frac{1}{2}\dot{\phi}^2+V(\phi)$, with $V(\phi)$ the
potential energy, then it follows that
$P_{\phi}+\rho_{\phi}=(1+\omega)\rho_{\phi}=\dot{\phi}^2$, and
hence $\dot{\phi}^2<0$ (i.e. classically an axionic component for
vacuum phantom energy [4]) if we want the weak energy condition to
be satisfied also for $\omega<-1$, which will in turn
automatically ensure violation of the dominant-energy condition].
No extra constraints are imposed to ensuring a causal propagation
condition that $dP/d\rho$ does not exceed unity.

I will argue however that whereas $P_{\phi}+\rho_{\phi} < 0$ or
$\phi\rightarrow i\Phi$ will suffice to ensure $\omega < -1$ and
phantom energy, or {\it vice versa}, in all quintessence cases
studied so far, the eventual emergence of a big rip in the future
will take only place in scalar-field models with equations of
state of the simplest form $P_{\phi}=\omega\rho_{\phi}$. In fact,
in case of a Chaplygin gas with equation of state
$P_{\phi}=-A/\rho_{\phi}^n$ (where $A$ and $n$ are positive
constants), which can in fact be regarded as just a particular
case of a dust fluid with a given bulk viscosity in a k=0
Friedmann-Robertson-Walker universe [5], the existence of phantom
energy does not lead to emergence of a big rip [6].

In this short letter I will show that a cosmological model with a
singular big rip at an arbitrary finite time in the future can be
also obtained when the scalar field satisfies equivalent
phantom-energy conditions in the case that it is equipped with
non-canonical kinetic energy for models restricted by a Lagrangian
of the form
\begin{equation}
{\it L}= K(\phi)q(x) ,
\end{equation}
where $x=\frac{1}{2}\Delta_{\mu}\phi\Delta_{\nu}\phi$. Such a
definition, which of course includes the quintessence model as a
limiting case, generally describes more general models claimed to
solve the coincidence problem without fine tuning, which have been
dubbed as k-essence [7]. Some of the current k-essence models
featured suitable tracking behaviour during radiation domination
with further attractors [7]. Introducing the usual variable
$y=1/\sqrt{x}$ and re-expressing $q(x)$ as $q[x(y)]\equiv g(y)/y$,
from the perfect-fluid analogy, we have for the pressure and
energy density of a generic k-essence scalar field $\phi$ [8]
\begin{equation}
P_{\phi}(y)=\frac{K(\phi)g(y)}{y}
\end{equation}
\begin{equation}
\rho_{\phi}(y)=-K(\phi)g'(y) ,
\end{equation}
where the prime means derivative with respect to $y$. Now, the
equation of state parameter and the effective sound speed can be
shown to be given by the $K(\phi)$-independent expressions
\begin{equation}
\omega_{\phi}(y)=-\frac{P_{\phi}}{\rho_{\phi}}=
-\frac{g(y)}{yg'(y)}
\end{equation}
\begin{equation}
c_{s\phi}^2(y)=\frac{P'_{\phi}}{\rho'_{\phi}}=
\frac{g(y)-yg'(y)}{y^2 g''(y)} .
\end{equation}
In general, k-essence models are defined by taking
$K(\phi)=\phi^{-2}>0$ [8]. Thus, for the weak energy condition to
hold it follows from Eq. (3) that in these models the function
$g(y)$ must be decreasing. Moreover, in these models it is
currently assumed that $c_{s\phi}^2 > 0$ and hence Eq. (5) implies
that $g''(y) > 0$, i.e. $g(y)$ should be a decreasing convex
function [8].

We set next the general form of the function $g(y)$ when we
consider a phantom-energy k-essence field; i.e. when we introduce
the following two phantom-energy conditions: $K(\phi)<0$ and
\[P_{\phi}(y)+\rho_{\phi}(y)\equiv 2K(\phi)xdq(x)/dx<0 ,\]
which are just the conditions that would follow, both at once,
whenever $\phi$ is made imaginary as in the quintessence models
[4]. However, since the kinetic term is non-canonical in the
k-essence scenario, the above two conditions should be defined by
themselves, not as being derived from the general formalism of
k-essence by Wick rotating the scalar field, for otherwise both
the variable $y$ and hence the function $g(y)$ would turn out to
be no longer real. In what follows we shall therefore introduce
the above two phantom-energy conditions while keeping $y$ and
$g(y)$ real.

Now, the first of these conditions and Eq. (3) amount to $g'(y)>0$
in order for satisfying the weak energy condition $\rho_{\phi}>0$,
and then from $g'(y)>0$ and the second phantom-energy condition,
we deduce that $g(y)>yg'(y)$. Whence using $g'(y)>0$, it also
follows that $g(y)>0$. Therefore the function $g(y)$ should be an
increasing concave function, that is we must also set $g''(y)<0$.
We have then from Eq. (5) that the square of the speed of sound
should necessarily be definite negative. However, even though an
imaginary sound speed would at first sight mean catastrophic
accelerated collapse of inhomogeneities, such a kind of
instability could still be avoided at least at the subhorizon
scale by taking into account the dependence of the sound speed on
the wavelength characterizing the instabilities [9]. Finally, it
is also a consequence from the above two phantom-energy conditions
that $\omega_{\phi}(y)<-1$.

\begin{figure}

\includegraphics[width=.9\columnwidth]{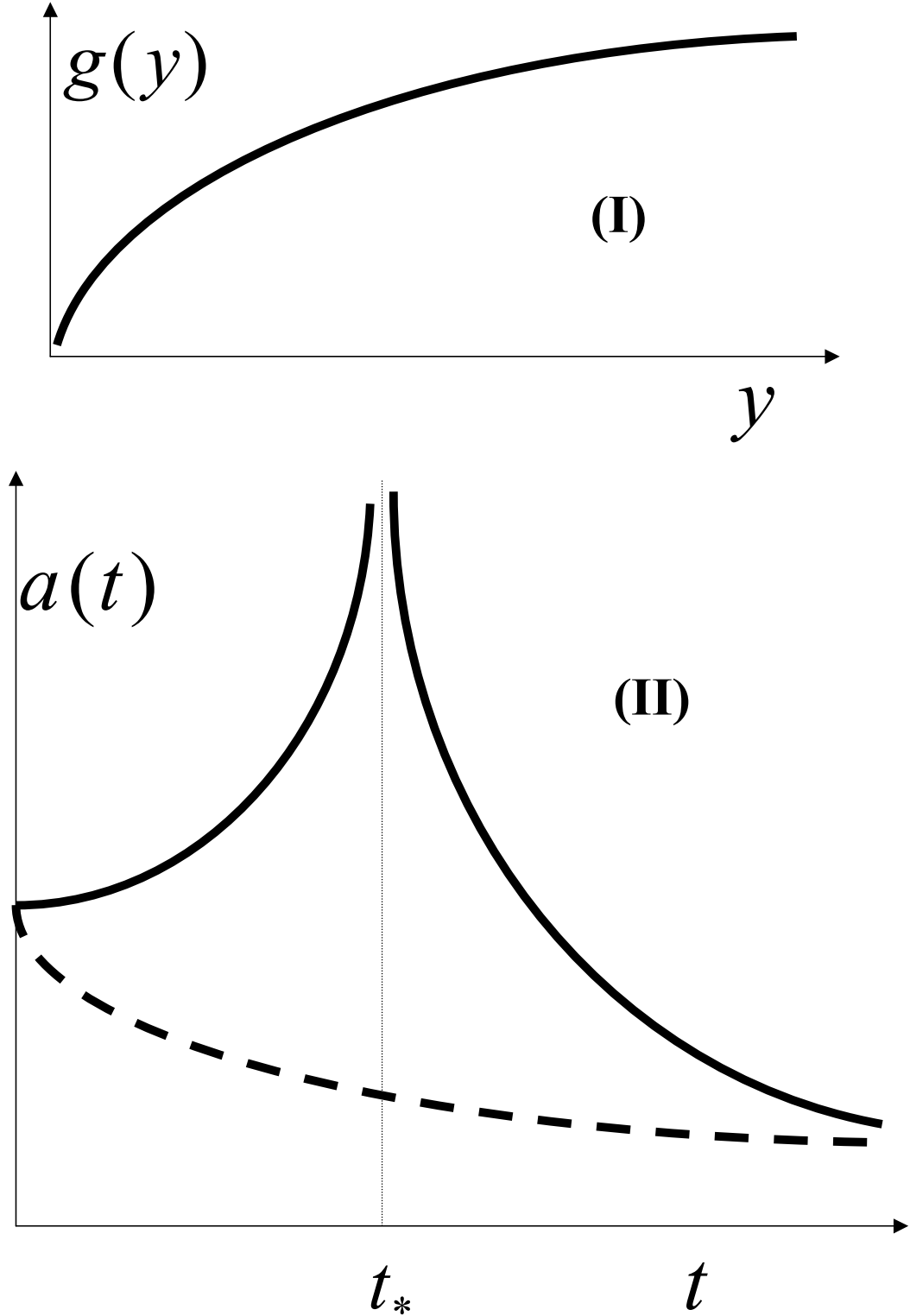}
\caption{\label{fig:epsart} (I) Generic shape of the function
$g(y)$ for a k-essence vacuum phantom-energy field. All units in
the plot are arbitrary. (II) Evolution of the scale factor $a(t)$
with cosmological time $t$ for a function $g(y)$ with the form
given in Fig. 1 (I). The dashed line for a universe with an ever
decreasing size corresponds to a choice of the sign of the
integration constant such that $t_*<0$. If we choose that sign to
be positive (solid line), then the constant $t_*$ becomes an
arbitrary time in the future at which the big rip takes place. All
units in the plot of Fig. 1 (II) are also arbitrary.}
\end{figure}

A simplest family of $g$-functions satisfying the above
requirements is (see Fig. 1 (I))
\begin{equation}
g(y)=By^{\beta} ,
\end{equation}
with $B$ and $\beta$ being given constants such that $B>0$ and
$0<\beta<1$. Actually a more general function $g(y)$ can be
written as a polynomials $g(y)=\sum_i B_i y^{\beta_i}$, where the
first term is given by Eq. (6) and all other extra terms are
characterized with powers $0<\beta_i <1$ as well and coefficients
$B>B_1 >B_2 >...$. It is moreover worth mentioning that the
polynomial $g(y)$ cases seem to be linked to the eight asymptotes
discussed by Barrow [10] when applying Fowler theorems for
first-order differential equations to obtain solutions of the
Raychaudhuri equation which are continuous, finite and monotonic
as $t\rightarrow\infty$. Even though rigorously checking whether
or not such a connection actually exist is outside the scope of
the present work, it would appear interesting to investigate it.
However, for the aims of this letter it will suffice taking only
the first term of such a polynomials. Let us specialize then in
the case of a spatially flat Friedmann-Robertson-Walker spacetime
with line element
\begin{equation}
ds^2=-dt^2+a(t)^2 d{\rm r}^2 ,
\end{equation}
in which $a(t)$ is the scale factor. In the case of a universe
dominated by a k-essence phantom vacuum energy, the Einstein field
equations are then
\begin{equation}
3H^2=\rho_{\phi}(y), \;\;\; 2\dot{H}+\rho_{\phi}(y)+P_{\phi}(y)=0
,
\end{equation}
with $H=\dot{a}/a$, the overhead dot meaning time derivative,
$\dot{ }=d/dt$. Combining the two expressions in Eq. (8) and using
the equation of state we can obtain for the function $g(y)$ as
given by Eq. (6)
\begin{equation}
3H^2=\frac{2\dot{H}\beta}{1-\beta} .
\end{equation}
For our spatially flat case we have then the solutions
\begin{equation}
a\propto\frac{1}{\left(t-t_*\right)^{2\beta/[3(1-\beta)]}} ,\;\;
0<\beta<1 ,
\end{equation}
where $t_*$ is an arbitrary constant. If we choose $t_*<0$, then
the scale factor would ever decrease with time (see Fig 1 (II),
dashed line). Obviously this solution family does not represent an
accelerating universe and should therefore be discarded. Of quite
greater interest is the choice $t_*>0$ for which the universe
(Fig. 1 (II), solid line) will first expand to reach a big-rip
singularity at the arbitrary time $t=t_*$ in the future, to
thereafter steadily collapse to zero at infinity; that is it
matches the behaviour expected for current quintessence models
with $\omega<-1$. The potentially dramatic difference is that
whereas in quintessence models the time at which the big rip will
occur depends nearly inversely on the absolute value of the state
equation parameter, in the present k-essence model the time $t_*$
is a rather arbitrary parameter.

In the case that we take for the field potential the usual
expression $K(\phi)=\phi^{-2}$ [8], the Euler-Lagrange equation
for the current k-essence field can also be written as
\begin{equation}
y^3\frac{d^2 g(y)}{dy^2}\ddot{\phi}
-3Hy\left[y\frac{dg(y)}{dy}-g(y)\right]\dot{\phi}+
\frac{4\frac{dg(y)}{dy}}{\phi}=0 .
\end{equation}
Therefore, using Eqs. (6) and (10) in the case that
$K(\phi)=\phi^{-2}$ one can integrate Eq. (11) to obtain for the
phantom-energy k-essence field
\begin{equation}
\phi=D_0\left[a_0^{3/\beta}\left(t-
t_*\right)^{\frac{\beta+1}{\beta-1}}
+E_0\right]^{\frac{\beta-1}{\beta+1}} ,
\end{equation}
with $D_0$ and $E_0$ being arbitrary integration constants and
$a_0$ an also arbitrary pre-factor for the scale factor in Eq.
(10). We notice that the phantom field $\phi$ tends to vanish as
$t\rightarrow t_*$, and hence its potential, $V=K(\phi)=
\phi^{-2}$, blows up at the big rip, such as it happens in
quintessence models [4]. It follows as well that the energy
density for the phantom field will increase initially as
$t\rightarrow t_*$ and blow up at $t=t_*$, as one should expect.

The main result in this letter is that phantom vacuum-energy leads
to a big rip singularity also for k-essence dark energy. Moreover,
one can play with the arbitrary values of the pre-factor $a_0$ for
the scale factor expression in Eq. (10) and those unboundedly
small positive values of $t_*$ which satisfy the observational
constraint [11] $1>\beta>0.7$ (note that in the present model
$\beta=-1/\omega_{\phi}$) and the currently observed cosmic
acceleration rate [11] to check that such a set of present
observations [11,12] is compatible with unboundedly small positive
values of $t_*$. Even though unboundedly larger values of $t_*$
are also allowed this way, k-essence phantom energy certainly may
well allow a very near occurrence of the big rip in the future.
Therefore, one could say that, in cosmological-time terms, a far
and a near occurrence of the big rip are similarly probable and
that, as a consequence from this, in the framework of cosmic
k-essence, cosmic doomsday might be awaiting us around the corner.

\vspace{.8cm}

\noindent{\bf Acknowledgements} The author thanks Carmen L.
Sig\"{u}enza, M. Bouhmadi and A. Jim\'{e}nez-Madrid for useful
discussions. This work was supported by DGICYT under Research
Project BMF2002-03758.

\pagebreak

\noindent\section*{References}

\begin{description}
\item [1] R.R. Caldwell, Phys. Lett. B545 (2002) 23; A.E. Schulz
and M.J. White, Phys. Rev. D64 (2001) 043514; J.G. Hao and X. Z.
Li, Phys. Rev. D67 (2003) 107303; G.W. Gibbons, hep-th/0302199; S.
Nojiri and S.D. Odintsov, Phys. Lett. B562 (2003) 147; B565 (2003)
1; B571 (2003) 1; P.Singh, M. Sami and N. Dadhich, Phys. Rev. D68
(2003) 023522; J.G. Hao and X.Z. Li, Phys. Rev. D68 (2003) 043501;
083514; M.P. Dabrowski, T. Stachowiak and M. Szydlowski, Phys.
Rev. D68 (2003) 067301; E. Elizalde and J. Quiroga H.,
gr-qc/0310128; V.B. Johri, astro-ph/0311293
\item [2] R.R. Caldwell, M. Kamionkowski and N.N. Weinberg, Phys. Rev.
Lett. 91 (2003) 071301.
\item [3] J.D. Barrow, G.J. Galloway and F.J. Tipler, Mon. Not.
Roy. Astr. Soc. 223 (1986) 835 .
\item [4] P.F. Gonz\'{a}lez-D\'{\i}az, {\it Axion Phantom Energy}, Report No.
IMAFF-RCA-03-08; L.P. Chimento and R. Lazkoz, Phys. Rev. Lett. 91
(2003) 211301.
\item [5] J.D. Barrow, Nucl. Phys. B310 (1988) 743 .
\item [6] P.F. Gonz\'{a}lez-D\'{\i}az, Phys. Rev. D68 (2003) 021303(R);
M. Bouhmadi and J.A. Jim\'{e}nez-Madrid, in preparation.
\item [7] C. Armend\'{a}riz-Pic\'{o}n, T. Damour and V. Mukhanov, Phys.
Lett. B458 (1999) 209; J. Garriga and V. Mukhanov, Phys. Lett.
B458 (1999) 219; T. Chiba, T. Okabe and M. Yamaguchi, Phys. Rev.
D62 (2000) 023511; C. Amend\'{a}riz-Pic\'{o}n, V. Mukhanov and P.J.
Steinhardt, Phys. Rev. Lett. 85 (2000) 4438; C. Amend\'{a}riz-Pic\'{o}n,
V. Mukhanov and P.J. Steinhardt, Phys. Rev. D63 (2001) 103510;
L.P. Chimento and A. Feinstein, astro-ph/0305007
\bibitem [8] M. Malquarti, E.J. Copeland and A.R. Liddle, Phys.
Rev. D68 (2003) 023512.
\item [9] P.J. Steinhardt, {\it Quintessential Cosmology and Cosmic
Acceleration}, http://feynman.princeton.edu/~steinh/prit4.ps
\item [10] J.D. Barrow, Class. Quant. Grav. 13 (1996) 2965 .
\item [11] D.N. Spergel {\it et al.}, Astrophys. J. Suppl. 148
(2003) 175. \item [12] S. Perlmutter {\it et al.}, Astrophys. J.
483 (1997) 565; S. Perlmutter {\it et al.}, Nature 391 (1998) 51;
P.M. Garnavich {\it et al.} Astrophys. J. Lett. 493 (1998) L53;
B.P. Schmidt, Astrophys. J. 507 (1998) 46; A.G. Riess {\it et al.}
Astrophys. J. 116 (1998) 1009; A. Riess {\it et al.}, Astrophys.
J. 560 (2001) 49; A.C. Baccigalupi, A. Balbi, S. Matarrase, F.
Perrotta and N. Vittorio, Phys. Rev. D65 (2002) 063520; M.
Melchiorri, L. Mersini, C.J. Odman and M. Tradden, Phys. Rev. D68
(2003) 043509; M. Doupis, A. Riazuelo, Y. Zolnierowski and A.
Blanchard, Astron. Astrophys. 405 (2003) 409.
\end{description}

\end{document}